\documentclass[%
 reprint,
nobibnotes,
 amsmath,amssymb,
 aps,
pra,
]{revtex4-2}

\usepackage{xcolor}
\usepackage{graphicx}
\usepackage{dcolumn}
\usepackage{bm}

\begin{document}


\title{Radiation damping of a Rayleigh scatterer illuminated by a plane wave}

\author{Mohammad Ali Abbassi}
\affiliation{Department of Electrical Engineering, Sharif University of Technology, Tehran, Iran}
\email[]{abbassi\_ma@yahoo.com}
 \author{Lukas Novotny}
\affiliation{Photonics Laboratory, ETH Zürich, CH-8093 Zürich, Switzerland}
\email[]{lnovotny@ethz.ch}




\date{\today}

\begin{abstract}
We investigate the radiation damping experienced by a dielectric spherical particle when it is illuminated by an electromagnetic plane wave within the Rayleigh regime. We derive the equivalent electric dipole of the moving particle and  subsequently calculate the electromagnetic force acting on it from two different approaches. In the first approach, we calculate the force from the integration of stress tensor and field momentum. In the second one, we calculate the force directly from the integration of the force density. Our derivations reveal that the damping coefficient is equal to $6P_{\mathrm{scat}}/mc^2$ along the propagation direction, whereas it is $P_{\mathrm{scat}}/mc^2$ along perpendicular directions. Here,  $P_{\mathrm{scat}}$ denotes the power scattered by the particle, and $mc^2$ represents the particle's mass energy.  The radiation damping derived in this study sets upper limits on the quality factor  of  optically levitated objects and ensures the existence of a steady-state solution of the particle's dynamics. 
\end{abstract}

\maketitle

\section{Introduction}
Arthur Ashkin, in his pioneering 1970 paper on trapping of particles by radiation pressure  writes ''The extension to vacuum of the present experiments on particle trapping in potential wells would be of interest since then any motions are frictionless" \cite{ashkin1970acceleration}. Later, in 1976, based on the Doppler effect he provides an estimate for the friction in vacuum (radiation damping) and concludes that the particle's oscillation will damp out with a half-time of roughly half a year \cite{ashkin1976optical}. Over the past decade,  optical trapping of levitated nanoparticles in high vacuum has gained renewed interest \cite{gonzalez2021levitodynamics,gieseler2018levitated,millen2020optomechanics},  and it has been shown that the particle's motion accelerates due to the random momentum transfer from photon scattering, so called photon recoil heating \cite{jain2016direct}.  Radiation damping was predicted to counteract this heating mechanism in order to establish a  stable equilibrium \cite{novotny2017radiation}.
\par
The rates of radiation damping and recoil heating are fundamental parameters in the study of optomechanics concerning levitated nanoparticles\cite{chang2010cavity,jain2016direct}. In his renowned 1905 paper on special relativity, Einstein calculated the radiation pressure acting on a moving totally reflecting mirror, employing the principle of energy conservation\cite{einstein1905electrodynamics}. From this calculation, the radiation damping rate for a movable mirror could be derived by linearizing the radiation pressure with respect to the particle's velocity. Further analysis of the friction forces due to the electromagnetic radiation for mirrors has been presented for one-dimensional structures like Fabry–Pérot cavities\cite{braginski1967ponderomotive,matsko1996value}. Moreover, radiation reaction forces of accelerated charges and two-level systems like atoms has been extensively studied in the literature\cite{hartemann1995classical,tamburini2010radiation,singal2016compatibility,wineland1979laser,wu2013radiation, horsley2011radiation,gould1997laser}. Radiation reaction of charged particles are usually studied by using retarded Liénard-Wiechert fields\cite{jackson2021classical}. When an atom is moving toward a red-detuned laser, its momentum decreases due to the Doppler effect\cite{gould1997laser}. In a paper by one of our authors\cite{novotny2017radiation}, the radiation damping of a polarizable particle interacting with an incident plane wave was derived by integrating the Maxwell stress tensor in the rest frame. However, we have found that this analysis is not complete, and that the contribution of the field momentum has to be accounted for when the integration is performed in the particle's rest frame. Moreover, the scattered far field of a moving particle is no longer transverse  in the laboratory frame, which affects the net momentum flux. In this paper, we revise the derivation for the radiation damping, taking into account the two mentioned corrections. We also present an alternative derivation by directly integrating the Lorentz force density. The results obtained from both approaches are in agreement.\par
In the following, we explore the scattering problem of a moving dielectric particle in the Rayleigh regime when it is illuminated by an incident plane wave in Sec. II. Then, we investigate the calculation of the force acting on the moving particle in Sec. III. We introduce two approaches for calculating the force, and consequently the radiation damping. The first involves surface integration of the stress tensor and field momentum in the particle's rest frame, presented in Sec. III A. The second method entails directly integrating the force density, presented in Sec. III B. Finally, the conclusions are made in Sec. IV.

\section{Dipole approximation for a moving particle}
Consider a dielectric spherical particle moving with velocity $\mathbf{v}$ and being illuminated by a monochromatic plane wave. The incident wave in the laboratory frame is considered as
\begin{subequations}
\label{eq:EHi}
\begin{equation}\mathbf{E}_i=\mathrm{Re}\Big[E_0\hat{\mathbf{x}} e^{i(kz-\omega t)}\Big],\end{equation}
\begin{equation}\mathbf{H}_i=\mathrm{Re}\Big[\frac{E_0}{\eta_0}\hat{\mathbf{y}} e^{i(kz-\omega t)}\Big],\end{equation}
\end{subequations}
where $\omega$ is the angular frequency of the incident wave, and $k=\omega/c$ is its wave number. Here, $\eta_0$ and $c$ denote the characteristic impedance and the speed of light in vacuum, respectively.\par
When $v\ll c$, the incident electric field in the rest frame of the particle can be expressed as\footnote{The rest frame of the particle is the inertial frame moving with velocity $\mathbf{v}$ with respect to the laboratory frame. We use prime notation to denote quantities in the rest frame.}
\begin{equation}\mathbf{E}_i^\prime=\mathrm{Re}\left[E_0\left(\hat{\mathbf{x}}+\bm{\beta}\times \hat{\mathbf{y}}\right)e^{ik(z^\prime-\bm{\beta}\cdot \mathbf{r}^\prime)}e^{-i\omega^\prime t^\prime}\right],\end{equation}
where $\mathbf{r}^\prime=\mathbf{r}-\mathbf{v}t$ and $t^\prime=t-\mathbf{v}\cdot \mathbf{r}/c^2$ are the space and time in that frame, and $\bm{\beta}=\mathbf{v}/c$ represents the normalized velocity of the particle. The incident wave in the rest frame is also a monochromatic plane wave with angular frequency of $\omega^\prime=\omega(1-\beta_z)$. In fact, we are dealing with a not moving spherical particle in the rest frame being illuminated by a monochromatic field $\mathbf{E}_i^\prime$. When the particle's radius is much smaller than the wavelength of the incident wave referred to as the Rayleigh regime, the particle acts as an electric dipole. The induced electric dipole is given by
\begin{equation}\label{eq:p}\begin{split}\mathbf{p}&=\mathrm{Re}\left[\underline{\mathbf{p}}e^{-i\omega^\prime t^\prime}\right]\\&=\mathrm{Re}\left[\alpha(\omega^\prime)E_0 \left(\hat{\mathbf{x}}+\bm{\beta}\times \hat{\mathbf{y}}\right)e^{ik(z^\prime_p-\bm{\beta}\cdot \mathbf{r}^\prime_p)} e^{-i\omega^\prime t^\prime}\right],\end{split}\end{equation}
where $\alpha$ denotes the polarizability of the particle. For a spherical particle with radius $R_p$ and dielectric constant $\epsilon_p$, the polarizability can be written as
\begin{equation}\label{eq:alpha}\alpha(\omega)=\frac{\alpha_0}{1-i\omega^3\alpha_0/6\pi\epsilon_0c^3},\end{equation}
with
$\alpha_0=4\pi\epsilon_0R_p^3(\epsilon_p-1)/(\epsilon_p+2)$
being the quasi-static polarizability of a sphere\cite{chaumet2000time,abbassi2018inclusion,abbassi2022self,abbassi2019green}.\par
In the rest frame, the electromagnetic fields scattered by the particle can be written as
\begin{subequations}
\begin{equation}\mathbf{E}_s^\prime=\mathrm{Re}\left[{\omega^\prime}^2\mu_0 \mathbf{G}_0(\mathbf{r}^\prime,\mathbf{r}_p^\prime;\omega^\prime)\cdot \underline{\mathbf{p}} e^{-i\omega^\prime t^\prime}\right],\end{equation}
\begin{equation}\mathbf{H}_s^\prime=\mathrm{Re}\left[-i\omega^\prime \bm{\nabla}\times \mathbf{G}_0(\mathbf{r}^\prime,\mathbf{r}_p^\prime;\omega^\prime)\cdot \underline{\mathbf{p}}e^{-i\omega^\prime t^\prime}\right].\end{equation}
\end{subequations}
Here, $\mathbf{G}_0$ represents the dyadic Green's function in the free space, given by
\begin{equation}\begin{split}\mathbf{G}_0(\mathbf{r}^\prime,\mathbf{r}_p^\prime;\omega^\prime)=\frac{e^{ik^\prime R^\prime}}{4\pi R^\prime}\Bigg[\left(1+\frac{i}{k^\prime R^\prime}-\frac{1}{{k^\prime}^2{R^\prime}^2}\right)&\mathbf{I}\\+\left(\frac{3}{{k^\prime}^2{R^\prime}^2}-\frac{3i}{k^\prime R^\prime}-1\right)&\frac{\mathbf{R}^\prime\mathbf{R}^\prime}{{R^\prime}^2}\Bigg],\end{split}\end{equation}
with $k^\prime=\omega^\prime/c$ being the wave number of the field in the rest frame, and $\mathbf{R}^\prime=\mathbf{r}^\prime-\mathbf{r}_p^\prime$\cite{novotny2012principles}. Without loss of generality, we assume that the particle is located at the origin in the rest frame, i.e., $\mathbf{r}_p^\prime=0$. Then, the scattered electromagnetic fields in the rest frame can be written as
\begin{widetext}
\begin{subequations}
\begin{equation}\mathbf{E}_s^\prime=\mathrm{Re}\left\{{\omega^\prime}^2\mu_0 \alpha(\omega^\prime)E_0 \frac{e^{i(k^\prime R^\prime-\omega^\prime t^\prime)}}{4\pi R^\prime}\begin{bmatrix}(1-\beta_z)(A_1+A_2\sin^2\theta^\prime \cos^2\phi^\prime)+\beta_xA_2 \sin\theta^\prime \cos\theta^\prime \cos\phi^\prime\\(1-\beta_z)A_2\sin^2\theta^\prime\cos\phi^\prime\sin\phi^\prime+ \beta_x A_2\sin\theta^\prime \cos\theta^\prime \sin\phi^\prime \\ \beta_xA_1+(1-\beta_z)A_2\sin\theta^\prime\cos\theta^\prime\cos\phi^\prime+\beta_x A_2\cos^2\theta^\prime\end{bmatrix}\right\},\end{equation}
\begin{equation}\mathbf{H}_s^\prime=\mathrm{Re}\left\{\frac{{\omega^\prime}^2\mu_0}{\eta_0}\alpha(\omega^\prime)E_0\frac{e^{i(k^\prime R^\prime-\omega^\prime t^\prime)}}{4\pi R^\prime}\begin{bmatrix} \beta_x A_3 \sin\theta^\prime \sin\phi^\prime \\(1-\beta_z)A_3\cos\theta^\prime -\beta_x A_3\sin\theta^\prime\cos\phi^\prime \\- (1-\beta_z)A_3\sin\theta^\prime \sin\phi^\prime\end{bmatrix}\right\},\end{equation}
\end{subequations}
\end{widetext}
Here, $(R^\prime,\theta^\prime,\phi^\prime)$ represents the spherical coordinate system in the rest frame of the particle which is considered as
\begin{subequations}
\begin{equation}x^\prime=R^\prime \sin\theta^\prime \cos\phi^\prime,\end{equation}
\begin{equation}y^\prime=R^\prime \sin\theta^\prime \sin\phi^\prime,\end{equation}
\begin{equation}z^\prime=R^\prime \cos\theta^\prime,\end{equation}
\end{subequations}
 and the coefficients $A_1$, $A_2$, and $A_3$ are defined by
\begin{subequations}
\begin{equation}A_1=1+\frac{i}{k^\prime R^\prime}-\frac{1}{{k^\prime}^2{R^\prime}^2},\end{equation}
\begin{equation}A_2=-1-\frac{3i}{k^\prime R^\prime}+\frac{3}{{k^\prime}^2{R^\prime}^2}.\end{equation}
\begin{equation}A_3=1+\frac{i}{k^\prime R^\prime}.\end{equation}
\end{subequations}
To get the far-fields, it suffices to substitute  $A_1,A_3\to 1$, and $A_2\to -1$.\par
Now, we can obtain the scattered fields in the laboratory frame from the following transformations\cite{zangwill2013modern}: 
\begin{subequations}
\begin{equation}\mathbf{E}_s=\mathbf{E}_s^\prime-\eta_0\bm{\beta}\times \mathbf{H}_s^\prime,\end{equation}
\begin{equation}\mathbf{H}_s=\mathbf{H}_s^\prime+\frac{1}{\eta_0}\bm{\beta}\times \mathbf{E}_s^\prime.\end{equation}
\end{subequations}
Hence, the scattered fields in the laboratory frame can be written as
\begin{subequations}
\label{eq:EHs}
\begin{equation}\mathbf{E}_s=\mathrm{Re}\left\{{\omega^\prime}^2\mu_0 \alpha(\omega^\prime)E_0 \frac{e^{i(k^\prime R^\prime-\omega^\prime t^\prime)}}{4\pi R^\prime}\mathbf{e_s}(\theta^\prime,\phi^\prime)\right\},\end{equation}
\begin{equation}\mathbf{H}_s=\mathrm{Re}\left\{\frac{{\omega^\prime}^2\mu_0}{\eta_0}\alpha(\omega^\prime)E_0\frac{e^{i(k^\prime R^\prime-\omega^\prime t^\prime)}}{4\pi R^\prime}\mathbf{h_s}(\theta^\prime,\phi^\prime)\right\},\end{equation}
\end{subequations}
where
\begin{widetext}
\begin{equation}\mathbf{e_s}(\theta^\prime,\phi^\prime)=\begin{bmatrix}(1-\beta_z)(A_1+A_2\sin^2\theta^\prime \cos^2\phi^\prime)+\beta_x A_2\sin\theta^\prime \cos\theta^\prime \cos\phi^\prime +\beta_y A_3 \sin\theta^\prime \sin\phi^\prime+\beta_z A_3\cos\theta^\prime \\(1-\beta_z)A_2\sin^2\theta^\prime\cos\phi^\prime\sin\phi^\prime+ \beta_x A_2\sin\theta^\prime \cos\theta^\prime \sin\phi^\prime -\beta_x A_3\sin\theta^\prime\sin\phi^\prime\\ \beta_xA_1+(1-\beta_z)A_2\sin\theta^\prime\cos\theta^\prime\cos\phi^\prime+\beta_x A_2\cos^2\theta^\prime-\beta_x A_3 \cos\theta^\prime\end{bmatrix},\end{equation}
and
\begin{equation}\mathbf{h_s}(\theta^\prime,\phi^\prime)=\begin{bmatrix} \beta_xA_3 \sin\theta^\prime \sin\phi^\prime +\beta_y A_2\sin\theta^\prime\cos\theta^\prime\cos\phi^\prime-\beta_z A_2 \sin^2\theta^\prime\cos\phi^\prime\sin\phi^\prime\\(1-\beta_z)A_3\cos\theta^\prime -\beta_x A_3\sin\theta^\prime\cos\phi^\prime +\beta_z(A_1+A_2\sin^2\theta^\prime\cos^2\phi^\prime)-\beta_x A_2 \sin\theta^\prime \cos\theta^\prime\cos\phi^\prime\\- (1-\beta_z)A_3\sin\theta^\prime \sin\phi^\prime+\beta_x A_2 \sin^2\theta^\prime \cos\phi^\prime \sin\phi^\prime-\beta_y(A_1+A_2\sin^2\theta^\prime\cos^2\phi^\prime)\end{bmatrix}.\end{equation}
\end{widetext}

\section{Force calculation}
According to the Lorentz force law, the electromagnetic force density exerted on a charge density $\rho$ and a current density $\mathbf{j}$ is given by
\begin{equation}\mathbf{f}=\rho\mathbf{E}+\mathbf{j}\times \mathbf{B}.\end{equation}
where $\rho$ and $j$ are the charge and current densities, respectively. We can also express the force density as
\begin{equation}\mathbf{f}=\bm{\nabla}\cdot \mathbf{T}-\frac{\partial \mathbf{g}}{\partial t},\end{equation}
in which
\begin{equation}\mathbf{T}=\epsilon_0 \mathbf{E}\mathbf{E}+\mu_0 \mathbf{H}\mathbf{H}-\frac{1}{2}\left(\epsilon_0 |\mathbf{E}|^2+\mu_0 |\mathbf{H}|^2\right)\mathbf{I},\end{equation}
is the stress tensor, and
\begin{equation}\mathbf{g}=\mu_0\epsilon_0 \mathbf{E}\times \mathbf{H},\end{equation}
is the momentum of the electromagnetic fields. When $v\ll c$, the Lorentz transformation implies 
$\bm{\nabla}=\bm{\nabla}^\prime-(\mathbf{v}/c^2)\partial_t$
and
$\partial_t=\partial_{t^\prime}-\mathbf{v}\cdot \bm{\nabla^\prime}$. Hence, we can also express the force density as
\begin{equation}\mathbf{f}=\mathbf{\nabla}^\prime\cdot \mathbf{T}+(\mathbf{v}\cdot \mathbf{\nabla}^\prime)\mathbf{g}-\frac{\partial}{\partial t^\prime}(\mathbf{g}+\frac{\mathbf{v}}{c^2}\cdot \mathbf{T}),\end{equation}
To obtain the total force acted on the particle, we should calculate $\int \mathbf{f}d^3r$. Since the Jacobian determinant is equivalent to
$|\mathbf{J}(\mathbf{r},\mathbf{r}^\prime)|=1+O(v^2)$,
the total force can also be calculated from $\int \mathbf{f}d^3r^\prime$. Therefore, the total force exerted upon the particle can be written as
\begin{equation}\mathbf{F}=\oint \mathbf{T}\cdot d\mathbf{s}^\prime+\oint \mathbf{g} (\mathbf{v}\cdot d\mathbf{s}^\prime)-\int  \frac{\partial }{\partial t^\prime}(\mathbf{g}+\frac{\mathbf{v}}{c^2}\cdot \mathbf{T})d^3r^\prime.\end{equation}
We can also apply a time-average to eliminate the high oscillating force terms. The time-averaged force can be obtained from
\begin{equation}\label{eq:F_ave}\langle \mathbf{F} \rangle = \oint \langle \mathbf{T} \rangle \cdot d\mathbf{s}^\prime+\oint \langle \mathbf{g}\rangle (\mathbf{v}\cdot d\mathbf{s}^\prime).\end{equation}
\par
We can calculate the force acted on the moving particle from two distinct approaches: calculating the surface integral of the stress tensor and the field momentum, or directly calculating the volume integral of the force density. In the following, we investigate these two approaches in detail, separately.

\subsection{Surface integration of stress tensor and field momentum}
The time-averaged force acting on the particle can be determined by calculating the surface integrals of stress tensor and field momentum according to Eq. \ref{eq:F_ave}. As seen in Sec. II, the total electromagnetic fields can be represented as the combination of incident and scattered fields. Consequently, both the stress tensor and field momentum can be decomposed into three parts: 
\begin{subequations}
\begin{equation}\mathbf{T}=\mathbf{T}_{ii}+\mathbf{T}_{ss}+\mathbf{T}_{is},\end{equation}
\begin{equation}\mathbf{g}=\mathbf{g}_{ii}+\mathbf{g}_{ss}+\mathbf{g}_{is}.\end{equation}
\end{subequations}
$\mathbf{T}_{ii}$ and $\mathbf{g}_{ii}$ are the components of the stress tensor and the field momentum that solely pertain on the incident fields that are defined as
\begin{subequations}
\begin{equation}\label{eq:Tii}\mathbf{T}_{ii}=\epsilon_0 \mathbf{E}_i\mathbf{E}_i+\mu_0 \mathbf{H}_i\mathbf{H}_i-\frac{1}{2}\left(\epsilon_0 |\mathbf{E}_i|^2+\mu_0 |\mathbf{H}_i|^2\right)\mathbf{I},\end{equation}
\begin{equation}\label{eq:gii}\mathbf{g}_{ii}=\mu_0\epsilon_0 \mathbf{E}_i\times \mathbf{H}_i .\end{equation}
\end{subequations}
$\mathbf{T}_{ss}$ and $\mathbf{g}_{ss}$ are the ones that only pertain on the scattered field that are described by
\begin{subequations}
\begin{equation}\label{eq:Tss}\mathbf{T}_{ss}=\epsilon_0 \mathbf{E}_s\mathbf{E}_s+\mu_0 \mathbf{H}_s\mathbf{H}_s-\frac{1}{2}\left(\epsilon_0 |\mathbf{E}_s|^2+\mu_0 |\mathbf{H}_s|^2\right)\mathbf{I},\end{equation}
\begin{equation}\label{eq:gss}\mathbf{g}_{ss}=\mu_0\epsilon_0 \mathbf{E}_s\times \mathbf{H}_s.\end{equation}
\end{subequations}
Eventually, $\mathbf{T}_{is}$ and $\mathbf{g}_{is}$ represent the mutual terms, defined as
\begin{subequations}
\begin{equation}\begin{split}\mathbf{T}_{is}=\epsilon_0 \mathbf{E}_i\mathbf{E}_s+\epsilon_0 \mathbf{E}_s\mathbf{E}_i+\mu_0 \mathbf{H}_i\mathbf{H}_s+\mu_0 \mathbf{H}_s\mathbf{H}_i\\-\left(\epsilon_0 \mathbf{E}_i\cdot\mathbf{E}_s+\mu_0 \mathbf{H}_i\cdot\mathbf{H}_s\right)\mathbf{I},\end{split}\end{equation}
\begin{equation}\label{eq:gis}\mathbf{g}_{is}=\mu_0\epsilon_0\left(\mathbf{E}_i\times \mathbf{H}_s+\mathbf{E}_s\times \mathbf{H}_i\right).\end{equation}
\end{subequations}
We have evaluated the surface integrals of the stress tensor components in Appendix A. The results are
\begin{equation}\oint \langle \mathbf{T}_{ii}\rangle \cdot d\mathbf{s}^\prime=0,\end{equation}
\begin{equation}\oint \langle \mathbf{T}_{ss}\rangle \cdot d\mathbf{s}^\prime=-\frac{\omega^4\mu_0^2\epsilon_0\alpha_0^2E_0^2}{60\pi c }\left(6v_x \hat{\mathbf{x}}+7v_y \hat{\mathbf{y}}+7v_z \hat{\mathbf{z}}\right),\end{equation}
\begin{equation}\label{Tis}\oint \langle \mathbf{T}_{is}\rangle \cdot d\mathbf{s}^\prime=\frac{\omega^4 \mu_0^2 \epsilon_0 \alpha_0^2E_0^2}{12\pi}\left(1-4\frac{v_z}{c}\right)\hat{\mathbf{z}}.\end{equation}
We can also evaluate the field momentum integrals. In Appendix B we have demonstrated that
\begin{equation}\oint \langle \mathbf{g}_{ii}\rangle  \left(d\mathbf{s}^\prime \cdot \mathbf{v}\right)=0,\end{equation}
\begin{equation}\oint \langle \mathbf{g}_{ss}\rangle  \left(d\mathbf{s}^\prime \cdot \mathbf{v}\right)=\frac{\omega^4\mu_0^2\epsilon_0\alpha_0^2E_0^2}{60\pi c}\left(v_x \hat{\mathbf{x}}+2v_y \hat{\mathbf{y}}+2v_z \hat{\mathbf{z}}\right),\end{equation}
\begin{equation}\oint \langle \mathbf{g}_{is}\rangle  \left(d\mathbf{s}^\prime \cdot \mathbf{v}\right)=-\frac{\omega^4\mu_0^2\epsilon_0\alpha_0^2E_0^2}{12\pi c}v_z \hat{\mathbf{z}}.\end{equation}
Therefore, the time-averaged force exerted upon the particle is
\begin{equation}\langle \mathbf{F}\rangle=\frac{P_{\mathrm{scat}}}{c}\hat{\mathbf{z}}-\frac{P_{\mathrm{scat}}}{c^2}\left(v_x \hat{\mathbf{x}}+v_y \hat{\mathbf{y}}+6v_z \hat{\mathbf{z}}\right)\end{equation}
where $P_{\mathrm{scat}}=k^4\alpha_0^2E_0^2/12\pi\epsilon_0$ is the power scattered by the particle. The first term in the above expression represents the radiation pressure force that acts on the particle along the propagation direction. The second one represents the radiation damping force experienced by the particle. Hence, the radiation damping tensor can be defined as
\begin{equation}\Gamma_{\mathrm{rad}}=\frac{P_{\mathrm{scat}}}{mc^2}\begin{bmatrix}1&0&0\\0&1&0\\0&0&6\end{bmatrix}.\end{equation}
\subsection{Volume integration of the force density} 
Here, we present another approach for deriving the radiation damping. We calculate the force acting on the moving particle directly from the volume integration of the force density. According to the Lorentz force law, the force exerted on a charge density $\rho$ and the current density $\mathbf{j}$ can be obtained from
\begin{equation}\mathbf{F}=\int \rho \mathbf{E}+\mathbf{j}\times \mathbf{B} d^3{r},\end{equation}
where the volume integral should be taken over the charge distribution.\par
In the rest frame, the induced dipole $\mathbf{p}$ is not moving and is located at position $\mathbf{r}_p^\prime$. Thus, the charge and current densities in that frame can be written as
\begin{subequations}
\begin{equation}\rho^\prime=-\bm{\nabla}^\prime \cdot \left[\mathbf{p} \delta (\mathbf{r}^\prime-\mathbf{r}^\prime_p)\right],\end{equation}
\begin{equation}\mathbf{j}^\prime=\frac{\partial \mathbf{p}}{\partial t^\prime} \delta(\mathbf{r}^\prime-\mathbf{r}^\prime_p),\end{equation}
\end{subequations}
Then, we can obtain the charge and current densities in the laboratory frame from
$\rho=\rho^\prime+\frac{\mathbf{v}}{c^2}\cdot \mathbf{j}^\prime$
and
$\mathbf{j}=\mathbf{j}^\prime+\rho^\prime\mathbf{v}$,
respectively. Hence, we can express $\rho$ and $\mathbf{j}$ as a function of laboratory-frame variables as
\begin{subequations}
\begin{equation}\label{eq:rho}\rho=-\bm{\nabla}\cdot \mathbf{p} \delta\left[\mathbf{r}-\mathbf{r}_p(t)\right]-\mathbf{p}\cdot \bm{\nabla}\delta \left[\mathbf{r}-\mathbf{r}_p(t)\right],\end{equation}
\begin{equation}\label{eq:j}\begin{split}\mathbf{j}=\Big[\frac{\partial \mathbf{p}}{\partial t}+\left(\mathbf{v}\cdot \bm{\nabla}\right)\mathbf{p}-\mathbf{v}\bm{\nabla}\cdot \mathbf{p}\Big] \delta\left[\mathbf{r}-\mathbf{r}_p(t)\right] \\-\mathbf{v}\bigg(\mathbf{p}\cdot\bm{\nabla}\delta\left[\mathbf{r}-\mathbf{r}_p(t)\right]\Big).\end{split}\end{equation}
\end{subequations}
Upon substituting $\rho$ and $\mathbf{j}$ from Eqs. \ref{eq:rho} and \ref{eq:j} into the force expression and computing the volume integral, one obtains
\begin{equation}\begin{split}\mathbf{F}=\Big[&\left(\mathbf{p}\cdot\bm{\nabla}\right)\mathbf{E}+\frac{\partial \mathbf{p}}{\partial t}\times \mathbf{B}\\&- \mathbf{B}\times\left(\mathbf{v}\cdot\bm{\nabla}\right)\mathbf{p}+\mathbf{v}\times\left(\mathbf{p}\cdot\bm{\nabla}\right)\mathbf{B}\Big]_{\mathbf{r}=\mathbf{r}_p(t)}.\end{split}\end{equation}
We can also express the force as a function of the rest frame variables:
\begin{equation}\label{eq:F_prime}\begin{split}\mathbf{F}=\Big[&\left(\mathbf{p}\cdot\bm{\nabla}^\prime\right)\mathbf{E}^\prime+\frac{\partial \mathbf{p}}{\partial t^\prime}\times \mathbf{B}^\prime\\ &+\frac{\partial \mathbf{p}}{\partial t^\prime}\times \left(\frac{\mathbf{v}}{c^2}\times \mathbf{E}^\prime\right)-\left(\mathbf{p}\cdot \frac{v}{c^2}\right)\frac{\partial \mathbf{E}^\prime}{\partial t^\prime}\Big]_{\mathbf{r}^\prime=\mathbf{r}^\prime_p},\end{split}\end{equation}
in which $\mathbf{E}^\prime=\mathbf{E}+\mathbf{v}\times \mathbf{B}$ and $\mathbf{B}^\prime=\mathbf{B}-(\mathbf{v}/c^2)\times \mathbf{E}$ represent the electric and magnetic fields in the rest frame, respectively.\par
As discussed in Sec. II, the electromagnetic fields in the particle's rest frame are monochromatic with angular frequency $\omega^\prime$. Hence, we can represent the fields as
\begin{subequations}
\begin{equation}\mathbf{E}^\prime=\mathrm{Re}\left[\underline{\mathbf{E}}^\prime e^{-i\omega (1-\beta_z)t^\prime}\right],\end{equation}
\begin{equation}\mathbf{B}^\prime=\mathrm{Re}\left[\underline{\mathbf{B}}^\prime e^{-i\omega (1-\beta_z)t^\prime}\right],\end{equation}
\end{subequations}
with $\underline{\mathbf{E}}^\prime$ and $\underline{\mathbf{B}}^\prime$ being the complex amplitude of the electric and magnetic fields, respectively. Upon substituting these expressions in Eq. \ref{eq:F_prime}, it can be easily shown that the time-averaged force acting on the particle can be written as
\begin{equation}\langle \mathbf{F}\rangle=\frac{1}{2}\mathrm{Re}\Big[\underline{p}_k^\ast\bm{\nabla}^\prime \underline{E}_k^{\prime}+i\omega(1-\beta_z)\left(\underline{\mathbf{p}}^\ast\cdot \underline{\mathbf{E}}^\prime\right)\frac{\mathbf{v}}{c^2}\Big]_{\mathbf{r}^\prime=\mathbf{r}_p^\prime}.\end{equation}
When calculating the force, it's important to account for the total electric field. We can decompose the electric field that needs to be incorporated into the force expression into two parts:
\begin{equation}\underline{\mathbf{E}}^\prime=\underline{\mathbf{E}}^\prime_{\mathrm{inc}}+\underline{\mathbf{E}}^\prime_{\mathrm{rad}}.\end{equation}
The first term is the incident electric field, given by
\begin{equation}\mathbf{E}^\prime_{\mathrm{inc}}=E_0\left(\hat{\mathbf{x}}+\bm{\beta}\times \hat{\mathbf{y}}\right)e^{ik(z^\prime-\bm{\beta}\cdot \bm{r}^\prime)}.\end{equation}
The second term is referred to as the radiation field, defined as
\begin{equation}\underline{\mathbf{E}}^\prime_{\mathrm{rad}}=\omega^2\mu_0 \mathrm{Im}\left[\mathbf{G}_0(\mathbf{r}^\prime,\mathbf{r}_p^\prime)\right]\cdot \underline{\mathbf{p}}.\end{equation}
It should be noted that the remaining part of the scattered field that is proportional to the real part of $\mathbf{G}_0$ is singular at the particle's position, and should not be considered when calculating the force.\par
Now, we can decompose $\langle \mathbf{F}\rangle$ into two parts:
\begin{equation}\langle \mathbf{F}\rangle=\langle \mathbf{F}_{\mathrm{inc}}\rangle+\langle \mathbf{F}_{\mathrm{rad}}\rangle.\end{equation}
 The first term represents the force that $\underline{\mathbf{E}}^\prime_{\mathrm{inc}}$ applies to the induced dipole, given by 
\begin{equation}\langle \mathbf{F}_{\mathrm{inc}}\rangle=\frac{kE_0^2}{2}(1-2\beta_z)\mathrm{Im}\left[\alpha\right]\hat{\mathbf{z}}-\frac{\omega kE_0^2}{2}\beta_z\mathrm{Im}\left[\frac{\partial \alpha}{\partial \omega}\right]\hat{\mathbf{z}}.\end{equation}
Upon using Eq. \ref{eq:alpha}, it can be easily shown that $\langle \mathbf{F}_{\mathrm{inc}}\rangle$ can be simplified to
\begin{equation}\langle \mathbf{F}_{\mathrm{inc}}\rangle=\frac{P_{\mathrm{scat}}}{c}\left(\hat{\mathbf{z}}-5\frac{v_z}{c}\hat{\mathbf{z}}\right),\end{equation}
 The second force term represented by $\langle \mathbf{F}_{\mathrm{rad}}\rangle$ is the force that $\underline{\mathbf{E}}^\prime_{\mathrm{rad}}$ applies to the induced dipole. By employing the following relations that can be easily demonstrated from the Taylor expansion of $\mathbf{G}_0$
\begin{subequations}
\begin{equation}\mathrm{Im}\left[\mathbf{G}_0(\mathbf{r}_p^\prime,\mathbf{r}_p^\prime)\right]=\frac{\omega}{6\pi c}\mathbf{I},\end{equation}
\begin{equation}\mathrm{Im}\left[\bm{\nabla}\mathbf{G}_0(\mathbf{r}_p^\prime,\mathbf{r}_p^\prime)\right]=0,\end{equation}
\end{subequations}
we can find that
\begin{equation}\langle \mathbf{F}_{\mathrm{rad}}\rangle=-\frac{P_{\mathrm{scat}}}{c^2}\mathbf{v}.\end{equation}
Therefore, the total time-averaged force acting on the particle is
\begin{equation}\langle \mathbf{F} \rangle= \frac{P_{\mathrm{scat}}}{c}\hat{\mathbf{z}}-\frac{P_{\mathrm{scat}}}{c^2}\left(v_x\hat{\mathbf{x}}+v_y\hat{\mathbf{y}}+6v_z\hat{\mathbf{z}}\right).\end{equation}
This result matches the one obtained in Sec. III A through the integration of the stress tensor and field momentum.
\section{Conclusions}
In summary, we have derived the radiation damping experienced by a moving dielectric particle exposed to an incident plane wave using two distinct approaches: one involves the surface integration of the stress tensor and field momentum, and the other involves the integration of the force density, directly. Our analysis has shown that the damping coefficient along the propagation direction is $6P_{\mathrm{scat}}/mc^2$, while it equates to $P_{\mathrm{scat}}/mc^2$ in perpendicular directions. We note that radiation damping is a necessary ingredient for the existence of a  steady state solution of the particle's dynamics. While zero-point field fluctuations heat the particle's motion via radiation pressure shot noise, radiation damping cools the motion and gives rise to a steady-state solution in which heating and cooling are balanced. Such an  equilibrium is the prerequisite  for Einstein's  famous fluctuation formula and the particle nature of radiation \cite{ritz1909gegenwartigen}.
\appendix
\section{Derivation of stress tensor integrals}
Here, we provide a detailed derivation of stress tensor integrals. As previously discussed in Sec. III, the stress tensor is decomposed into three components. One of these components is $\mathbf{T}_{ii}$ that represents the stress tensor of the incident wave, as defined in Eq. \ref{eq:Tii}. Substituting the incident fields from Eqs. \ref{eq:EHi} results in
\begin{equation}\begin{split}\oint \langle \mathbf{T}_{ii}\rangle \cdot d\mathbf{s}^\prime=\oint \frac{1}{2}\mathrm{Re}\Bigg\{\epsilon_0 E_0^2\Big[ (\mathbf{e}_i^\ast \cdot \hat{\mathbf{R}}^\prime)\mathbf{e}_i+(\mathbf{h}_i^\ast \cdot \hat{\mathbf{R}}^\prime)\mathbf{h}_i\\-\frac{1}{2}\left(\mathbf{e}_i\cdot \mathbf{e}_i^\ast+\mathbf{h}_i\cdot \mathbf{h}_i^\ast\right)\hat{\mathbf{R}}^\prime\Big]\Bigg\}ds^\prime,\end{split}\end{equation}
Here, $\mathbf{e}_i=(1,0,0)$ and $\mathbf{h}_i=(0,1,0)$ represents the direction of the incident electric and magnetic fields, respectively, and $\hat{\mathbf{R}}^\prime=(\sin\theta^\prime \cos\phi^\prime,\sin\theta^\prime \sin\phi^\prime,\cos\theta^\prime)$. After evaluating the integration over the polar angle, i.e. $\phi^\prime$, we obtain 
\begin{equation}\oint \langle \mathbf{T}_{ii}\rangle \cdot d\mathbf{s}^\prime =\pi\epsilon_0E_0^2\int_0^\pi \begin{bmatrix}0\\0\\-\cos\theta^\prime\end{bmatrix}{R^\prime}^2 \sin\theta^\prime d\theta^\prime.\end{equation}
It can be easily shown that the above integral vanishes. Consequently,
\begin{equation}\oint \langle \mathbf{T}_{ii}\rangle \cdot d\mathbf{s}^\prime =0.\end{equation}
\par
Now, we aim calculating the surface integral of $\langle\mathbf{T}_{ss}\rangle$. As defined in Eq. \ref{eq:Tss}, $\mathbf{T}_{ss}$ is the component of the stress tensor associated solely with the scattered fields. Substituting the incident fields from Eqs. \ref{eq:EHs} yields
\begin{widetext}
\begin{equation}\oint \langle \mathbf{T}_{ss}\rangle \cdot d\mathbf{s}^\prime=\oint \frac{1}{2}\mathrm{Re}\Bigg\{\frac{{\omega^\prime}^4\mu_0^2\epsilon_0 \alpha_0^2 E_0^2}{16\pi^2 {R^\prime}^2}\Big[ (\mathbf{e}_s^\ast \cdot \hat{\mathbf{R}}^\prime)\mathbf{e}_s+(\mathbf{h}_s^\ast \cdot \hat{\mathbf{R}}^\prime)\mathbf{h}_s-\frac{1}{2}\left(\mathbf{e}_s\cdot \mathbf{e}_s^\ast+\mathbf{h}_s\cdot \mathbf{h}_s^\ast\right)\hat{\mathbf{R}}^\prime\Big]\Bigg\}ds^\prime.\end{equation}
We can use the far-fields values when calculating the above expression. After evaluating the integration over the polar angle, i.e. $\phi^\prime$, it becomes
\begin{equation}\oint \langle \mathbf{T}_{ss}\rangle \cdot d\mathbf{s}^\prime=\frac{{\omega^\prime}^4\mu_0^2\epsilon_0 \alpha_0^2 E_0^2}{32\pi}\int_0^\pi \begin{bmatrix}\beta_x \left(-\frac{5}{4}+2\cos\theta^\prime-\frac{3}{2}\cos^2 \theta^\prime-2\cos^3\theta^\prime+\frac{3}{4}\cos^4\theta^\prime\right)\\\beta_y\left(-\frac{7}{4}-\frac{1}{2}\cos^2\theta^\prime+\frac{1}{4}\cos^4\theta^\prime\right)\\-2\cos\theta^\prime-2\cos^3\theta^\prime +\beta_z\left(-1+2\cos\theta^\prime-2\cos^2\theta^\prime+2\cos^3\theta^\prime-\cos^4\theta^\prime\right)\end{bmatrix}\sin\theta^\prime d\theta^\prime.\end{equation}

If we evaluate the integral above and retain terms up to first order in $\beta$, one obtains
\begin{equation}\oint \langle \mathbf{T}_{ss}\rangle \cdot d\mathbf{s}^\prime=-\frac{\omega^4\mu_0^2\epsilon_0\alpha_0^2E_0^2}{60\pi c }\left(6v_x \hat{\mathbf{x}}+7v_y \hat{\mathbf{y}}+7v_z \hat{\mathbf{z}}\right).\end{equation}
\par
Eventually, we want to calculate the surface integral of the mutual component of the stress tensor, as defined in Eq. \ref{Tis}. Upon substituting the incident and scattered fields from Eqs. \ref{eq:EHi} and \ref{eq:EHs}, respectively, we obtain

\begin{equation}\begin{split}\oint \langle \mathbf{T}_{is}\rangle \cdot d\mathbf{s}^\prime=\oint \frac{1}{2}\mathrm{Re}\Bigg\{{\omega^\prime}^2\mu_0\epsilon_0 \alpha(\omega^\prime) E_0^2 \frac{e^{i(k^\prime R^\prime-\omega^\prime t^\prime-kz+\omega t)}}{4\pi R^\prime}\Big[ (\mathbf{e}_i^\ast \cdot \hat{\mathbf{R}}^\prime)\mathbf{e}_s+(\mathbf{e}_s \cdot \hat{\mathbf{R}}^\prime)\mathbf{e}_i^\ast+(\mathbf{h}_i^\ast \cdot \hat{\mathbf{R}}^\prime)\mathbf{h}_s+(\mathbf{h}_s \cdot \hat{\mathbf{R}}^\prime)\mathbf{h}_i^\ast\\-\left(\mathbf{e}_i^\ast\cdot \mathbf{e}_s+\mathbf{h}_i^\ast\cdot \mathbf{h}_s\right)\hat{\mathbf{R}}^\prime&\Big]\Bigg\}ds^\prime.\end{split}\end{equation}
If we retain terms up to first order in $\beta$ and compute the integral  over $\phi^\prime$, the expression becomes
\begin{equation}\begin{split}\oint \langle \mathbf{T}_{is}\rangle \cdot d\mathbf{s}^\prime=\mathrm{Re}\left[\frac{k^\prime \alpha(\omega^\prime)E_0^2}{8}\lim_{k^\prime R^\prime \to 0}\left\{k^\prime R^\prime \int_0^\pi e^{ik^\prime R^\prime (1-\cos \theta^\prime)}\begin{bmatrix} C_0^x+C_1^x\cos\theta^\prime+ C_2^x\cos^2\theta^\prime+C_3^x\cos^3\theta^\prime \\ C_0^y+C_1^y\cos\theta^\prime+ C_2^y\cos^2\theta^\prime \\ C_0^z+C_1^z\cos\theta^\prime+ C_2^z\cos^2\theta^\prime+C_3^z\cos^3\theta^\prime\end{bmatrix}\sin\theta^\prime d\theta^\prime\right\}\right].\end{split}\end{equation}
\end{widetext}
The coefficients appearing in the x-component of the above expression are defined as
\begin{subequations}
\begin{equation}C_0^x=\beta_x\left[A_3+ik^\prime R^\prime(A_1+A_2)\right],\end{equation}
\begin{equation}C_1^x=\beta_x\left(2A_1+3A_2-ik^\prime R^\prime A_3\right), \end{equation}
\begin{equation}C_2^x=\beta_x\left[-3A_3-ik^\prime R^\prime (A_1+A_2)\right],\end{equation}
\begin{equation}C_3^x=\beta_x\left(-A_2+ik^\prime R^\prime A_3\right) .\end{equation}
\end{subequations}
The ones appeared in the y-component are given by
\begin{subequations}
\begin{equation}C_0^y=\beta_y\left(-A_3-ik^\prime R^\prime A_1\right),\end{equation}
\begin{equation}C_1^y=-2\beta_y A_1,\end{equation}
\begin{equation}C_2^y=\beta_x\left(A_3+ik^\prime R^\prime A_1\right),\end{equation}
\end{subequations}
and eventually the ones in the z-component are
\begin{subequations}
\begin{equation}C_0^z=-(1-\beta_z)A_3,\end{equation}
\begin{equation}C_1^z=-2A_1-\beta_z A_2,\end{equation}
\begin{equation}C_2^z=-(1+\beta_z)A_3,\end{equation}
\begin{equation}C_3^z=\beta_z A_2.\end{equation}
\end{subequations}
Upon performing the integral and taking the limit, the resultant expression is:
\begin{equation}\oint \langle \mathbf{T}_{is}\rangle \cdot d\mathbf{s}^\prime=\frac{\omega^4 \mu_0^2 \epsilon_0 \alpha_0^2E_0^2}{12\pi}\left(1-4\frac{v_z}{c}\right)\hat{\mathbf{z}}.\end{equation}
\section{Derivation of field momentum integrals}
Here, we provide a detailed derivation of the field momentum integrals appeared in Eq. \ref{eq:F_ave}. As discussed previously, the field momentum can be decomposed into there components: $\mathbf{g}_{ii}$, $\mathbf{g}_{ss}$, and $\mathbf{g}_{is}$. According to the definition of $\mathbf{g}_{ii}$ given in Eq. \ref{eq:gii}, and upon substituting the incident fields from Eqs. \ref{eq:EHi}, one obtains
\begin{equation}\oint \langle \mathbf{g}_{ii}\rangle ( d\mathbf{s}^\prime\cdot \mathbf{v})=\oint \frac{1}{2}\mathrm{Re}\left[\frac{\epsilon_0\mu_0 E_0^2}{\eta_0} \hat{\mathbf{e}}_i\times\hat{\mathbf{h}}_i^\ast\right]( d\mathbf{s}^\prime\cdot \mathbf{v}),\end{equation}
\begin{widetext}
which can be expressed as:
\begin{equation}\oint \langle \mathbf{g}_{ii}\rangle ( d\mathbf{s}^\prime\cdot \mathbf{v})=\mathrm{Re}\Bigg\{\int_0^\pi \int_0^{2\pi}\frac{\epsilon_0E_0^2}{2}\begin{bmatrix}0\\0\\1\end{bmatrix} (\beta_x\sin\theta^\prime\cos\phi^\prime+\beta_y\sin\theta^\prime\sin\phi^\prime+\beta_z \cos\theta^\prime){R^\prime}^2\sin\theta^\prime d\theta^\prime d\phi^\prime\Bigg\}.\end{equation}
Upon computing the above integral, we can easily demonstrate that
\begin{equation}\oint \langle \mathbf{g}_{ii}\rangle ( d\mathbf{s}^\prime\cdot \mathbf{v})=0.\end{equation}
\par
Now, we aim the calculation of $\langle\mathbf{g}_{ss}\rangle$ integral. According to the definition of $\mathbf{g}_{ss}$ given in Eq. \ref{eq:gss}, and substituting the scattered fields from Eqs. \ref{eq:EHs}, we obtain 
\begin{equation}\oint \langle \mathbf{g}_{ss}\rangle ( d\mathbf{s}^\prime\cdot \mathbf{v})=\oint \frac{1}{2}\mathrm{Re}\Bigg\{\frac{{\omega^\prime}^4\mu_0^2\epsilon_0 \alpha_0^2 E_0^2}{16\pi^2 {R^\prime}^2}\hat{\mathbf{e}}_s\times \hat{\mathbf{h}}_s^\ast\Bigg\}( d\mathbf{s}^\prime\cdot \mathbf{v}),\end{equation}
that can be further expanded into
\begin{equation}\begin{split}\oint \langle \mathbf{g}_{ss}\rangle ( d\mathbf{s}^\prime\cdot \mathbf{v})=\frac{1}{2}\mathrm{Re}\Bigg\{\int_0^\pi \int_0^{2\pi} \frac{{\omega}^4\mu_0^2\epsilon_0 \alpha_0^2 E_0^2}{16\pi^2 {R^\prime}^2} \begin{bmatrix}\sin\theta^\prime\cos^2\theta^\prime\cos\phi^\prime+\sin^3\theta^\prime\sin^2\phi^\prime\cos\phi^\prime \\ \sin\theta^\prime\sin\phi^\prime-\sin^3\theta^\prime\sin\phi^\prime\cos^2\phi^\prime \\ \cos\theta^\prime-\sin^2\theta^\prime\cos\theta^\prime\cos^2\phi^\prime\end{bmatrix}\\ \times (\beta_x\sin\theta^\prime\cos\phi^\prime+\beta_y\sin\theta^\prime\sin\phi^\prime+\beta_z \cos\theta^\prime)&{R^\prime}^2\sin\theta^\prime d\theta^\prime d\phi^\prime\Bigg\}.\end{split}\end{equation}
If we evaluate the integral above and retain terms up to first order in $\beta$, one obtains
\begin{equation}\oint \langle \mathbf{g}_{ss}\rangle  \left(d\mathbf{s}^\prime \cdot \mathbf{v}\right)=\frac{\omega^4\mu_0^2\epsilon_0\alpha_0^2E_0^2}{60\pi c}\left(v_x \hat{\mathbf{x}}+2v_y \hat{\mathbf{y}}+2v_z \hat{\mathbf{z}}\right).\end{equation}
\par
Eventually, we aim the calculation of the mutual term. According the definition of $\mathbf{g}_{is}$ given in Eq. \ref{eq:gis}, it can be easily shown that
\begin{equation}\oint \langle \mathbf{g}_{is}\rangle  \left(d\mathbf{s}^\prime \cdot \mathbf{v}\right)=\oint \frac{1}{2}\mathrm{Re}\Bigg\{{\omega^\prime}^2\mu_0\epsilon_0 \alpha(\omega^\prime) E_0^2 \frac{e^{i(k^\prime R^\prime-\omega^\prime t^\prime-kz+\omega t)}}{4\pi R^\prime}\big[\hat{\mathbf{e}}_i\times \hat{\mathbf{h}}_s^\ast+\hat{\mathbf{e}}_s \times \hat{\mathbf{h}}_i^\ast\big]\Bigg\}( d\mathbf{s}^\prime\cdot \mathbf{v}).\end{equation}
Since we want to calculate the above expression up to first order terms in $\beta$, we can retain $\hat{\mathbf{e}}_i\times \hat{\mathbf{h}}_s^\ast+\hat{\mathbf{e}}_s \times \hat{\mathbf{h}}_i^\ast$ up to zero order in beta. Then, the above expression can be written as
\begin{equation}\begin{split}\oint \langle \mathbf{g}_{is}\rangle  \left(d\mathbf{s}^\prime \cdot \mathbf{v}\right)= \mathrm{Re}\Bigg[\frac{k^\prime \alpha(\omega^\prime)E_0^2}{8\pi}\lim_{k^\prime R^\prime \to 0} \Bigg\{k^\prime R^\prime \int_0^\pi \int_0^{2\pi} e^{ik^\prime R^\prime(1-\cos\theta^\prime)} \begin{bmatrix}A_2 \sin\theta^\prime \cos\theta^\prime \cos\phi^\prime \\ A_3\sin\theta^\prime\sin\phi^\prime \\ A_3\cos\theta^\prime-A_1-A_2 \sin^2\theta^\prime\cos^2\phi^\prime\end{bmatrix}\\\times(\beta_x\sin\theta^\prime\cos\phi^\prime+\beta_y\sin\theta^\prime\sin\phi^\prime+\beta_z \cos\theta^\prime)\sin\theta^\prime d\theta^\prime&  \Bigg\}\Bigg].\end{split}\end{equation}

Upon evaluating the above expression and retaining terms up to first order in $\beta$, one obtains 
\begin{equation}\oint \langle \mathbf{g}_{is}\rangle  \left(d\mathbf{s}^\prime \cdot \mathbf{v}\right)=-\frac{\omega^4\mu_0^2\epsilon_0\alpha_0^2E_0^2}{12\pi c}v_z \hat{\mathbf{z}}.\end{equation}
\end{widetext}



\begin{thebibliography}{27}%
\makeatletter
\providecommand \@ifxundefined [1]{%
 \@ifx{#1\undefined}
}%
\providecommand \@ifnum [1]{%
 \ifnum #1\expandafter \@firstoftwo
 \else \expandafter \@secondoftwo
 \fi
}%
\providecommand \@ifx [1]{%
 \ifx #1\expandafter \@firstoftwo
 \else \expandafter \@secondoftwo
 \fi
}%
\providecommand \natexlab [1]{#1}%
\providecommand \enquote  [1]{``#1''}%
\providecommand \bibnamefont  [1]{#1}%
\providecommand \bibfnamefont [1]{#1}%
\providecommand \citenamefont [1]{#1}%
\providecommand \href@noop [0]{\@secondoftwo}%
\providecommand \href [0]{\begingroup \@sanitize@url \@href}%
\providecommand \@href[1]{\@@startlink{#1}\@@href}%
\providecommand \@@href[1]{\endgroup#1\@@endlink}%
\providecommand \@sanitize@url [0]{\catcode `\\12\catcode `\$12\catcode
  `\&12\catcode `\#12\catcode `\^12\catcode `\_12\catcode `\%12\relax}%
\providecommand \@@startlink[1]{}%
\providecommand \@@endlink[0]{}%
\providecommand \url  [0]{\begingroup\@sanitize@url \@url }%
\providecommand \@url [1]{\endgroup\@href {#1}{\urlprefix }}%
\providecommand \urlprefix  [0]{URL }%
\providecommand \Eprint [0]{\href }%
\providecommand \doibase [0]{https://doi.org/}%
\providecommand \selectlanguage [0]{\@gobble}%
\providecommand \bibinfo  [0]{\@secondoftwo}%
\providecommand \bibfield  [0]{\@secondoftwo}%
\providecommand \translation [1]{[#1]}%
\providecommand \BibitemOpen [0]{}%
\providecommand \bibitemStop [0]{}%
\providecommand \bibitemNoStop [0]{.\EOS\space}%
\providecommand \EOS [0]{\spacefactor3000\relax}%
\providecommand \BibitemShut  [1]{\csname bibitem#1\endcsname}%
\let\auto@bib@innerbib\@empty
\bibitem [{\citenamefont {Ashkin}(1970)}]{ashkin1970acceleration}%
  \BibitemOpen
  \bibfield  {author} {\bibinfo {author} {\bibfnamefont {A.}~\bibnamefont
  {Ashkin}},\ }\bibfield  {title} {\bibinfo {title} {Acceleration and trapping
  of particles by radiation pressure},\ }\href@noop {} {\bibfield  {journal}
  {\bibinfo  {journal} {Physical review letters}\ }\textbf {\bibinfo {volume}
  {24}},\ \bibinfo {pages} {156} (\bibinfo {year} {1970})}\BibitemShut
  {NoStop}%
\bibitem [{\citenamefont {Ashkin}\ and\ \citenamefont
  {Dziedzic}(1976)}]{ashkin1976optical}%
  \BibitemOpen
  \bibfield  {author} {\bibinfo {author} {\bibfnamefont {A.}~\bibnamefont
  {Ashkin}}\ and\ \bibinfo {author} {\bibfnamefont {J.}~\bibnamefont
  {Dziedzic}},\ }\bibfield  {title} {\bibinfo {title} {Optical levitation in
  high vacuum},\ }\href@noop {} {\bibfield  {journal} {\bibinfo  {journal}
  {Applied Physics Letters}\ }\textbf {\bibinfo {volume} {28}},\ \bibinfo
  {pages} {333} (\bibinfo {year} {1976})}\BibitemShut {NoStop}%
\bibitem [{\citenamefont {Gonzalez-Ballestero}\ \emph
  {et~al.}(2021)\citenamefont {Gonzalez-Ballestero}, \citenamefont
  {Aspelmeyer}, \citenamefont {Novotny}, \citenamefont {Quidant},\ and\
  \citenamefont {Romero-Isart}}]{gonzalez2021levitodynamics}%
  \BibitemOpen
  \bibfield  {author} {\bibinfo {author} {\bibfnamefont {C.}~\bibnamefont
  {Gonzalez-Ballestero}}, \bibinfo {author} {\bibfnamefont {M.}~\bibnamefont
  {Aspelmeyer}}, \bibinfo {author} {\bibfnamefont {L.}~\bibnamefont {Novotny}},
  \bibinfo {author} {\bibfnamefont {R.}~\bibnamefont {Quidant}},\ and\ \bibinfo
  {author} {\bibfnamefont {O.}~\bibnamefont {Romero-Isart}},\ }\bibfield
  {title} {\bibinfo {title} {Levitodynamics: Levitation and control of
  microscopic objects in vacuum},\ }\href@noop {} {\bibfield  {journal}
  {\bibinfo  {journal} {Science}\ }\textbf {\bibinfo {volume} {374}},\ \bibinfo
  {pages} {eabg3027} (\bibinfo {year} {2021})}\BibitemShut {NoStop}%
\bibitem [{\citenamefont {Gieseler}\ and\ \citenamefont
  {Millen}(2018)}]{gieseler2018levitated}%
  \BibitemOpen
  \bibfield  {author} {\bibinfo {author} {\bibfnamefont {J.}~\bibnamefont
  {Gieseler}}\ and\ \bibinfo {author} {\bibfnamefont {J.}~\bibnamefont
  {Millen}},\ }\bibfield  {title} {\bibinfo {title} {Levitated nanoparticles
  for microscopic thermodynamics—a review},\ }\href@noop {} {\bibfield
  {journal} {\bibinfo  {journal} {Entropy}\ }\textbf {\bibinfo {volume} {20}},\
  \bibinfo {pages} {326} (\bibinfo {year} {2018})}\BibitemShut {NoStop}%
\bibitem [{\citenamefont {Millen}\ \emph {et~al.}(2020)\citenamefont {Millen},
  \citenamefont {Monteiro}, \citenamefont {Pettit},\ and\ \citenamefont
  {Vamivakas}}]{millen2020optomechanics}%
  \BibitemOpen
  \bibfield  {author} {\bibinfo {author} {\bibfnamefont {J.}~\bibnamefont
  {Millen}}, \bibinfo {author} {\bibfnamefont {T.~S.}\ \bibnamefont
  {Monteiro}}, \bibinfo {author} {\bibfnamefont {R.}~\bibnamefont {Pettit}},\
  and\ \bibinfo {author} {\bibfnamefont {A.~N.}\ \bibnamefont {Vamivakas}},\
  }\bibfield  {title} {\bibinfo {title} {Optomechanics with levitated
  particles},\ }\href@noop {} {\bibfield  {journal} {\bibinfo  {journal}
  {Reports on Progress in Physics}\ }\textbf {\bibinfo {volume} {83}},\
  \bibinfo {pages} {026401} (\bibinfo {year} {2020})}\BibitemShut {NoStop}%
\bibitem [{\citenamefont {Jain}\ \emph {et~al.}(2016)\citenamefont {Jain},
  \citenamefont {Gieseler}, \citenamefont {Moritz}, \citenamefont {Dellago},
  \citenamefont {Quidant},\ and\ \citenamefont {Novotny}}]{jain2016direct}%
  \BibitemOpen
  \bibfield  {author} {\bibinfo {author} {\bibfnamefont {V.}~\bibnamefont
  {Jain}}, \bibinfo {author} {\bibfnamefont {J.}~\bibnamefont {Gieseler}},
  \bibinfo {author} {\bibfnamefont {C.}~\bibnamefont {Moritz}}, \bibinfo
  {author} {\bibfnamefont {C.}~\bibnamefont {Dellago}}, \bibinfo {author}
  {\bibfnamefont {R.}~\bibnamefont {Quidant}},\ and\ \bibinfo {author}
  {\bibfnamefont {L.}~\bibnamefont {Novotny}},\ }\bibfield  {title} {\bibinfo
  {title} {Direct measurement of photon recoil from a levitated nanoparticle},\
  }\href@noop {} {\bibfield  {journal} {\bibinfo  {journal} {Physical review
  letters}\ }\textbf {\bibinfo {volume} {116}},\ \bibinfo {pages} {243601}
  (\bibinfo {year} {2016})}\BibitemShut {NoStop}%
\bibitem [{\citenamefont {Novotny}(2017)}]{novotny2017radiation}%
  \BibitemOpen
  \bibfield  {author} {\bibinfo {author} {\bibfnamefont {L.}~\bibnamefont
  {Novotny}},\ }\bibfield  {title} {\bibinfo {title} {Radiation damping of a
  polarizable particle},\ }\href@noop {} {\bibfield  {journal} {\bibinfo
  {journal} {Physical Review A}\ }\textbf {\bibinfo {volume} {96}},\ \bibinfo
  {pages} {032108} (\bibinfo {year} {2017})}\BibitemShut {NoStop}%
\bibitem [{\citenamefont {Chang}\ \emph {et~al.}(2010)\citenamefont {Chang},
  \citenamefont {Regal}, \citenamefont {Papp}, \citenamefont {Wilson},
  \citenamefont {Ye}, \citenamefont {Painter}, \citenamefont {Kimble},\ and\
  \citenamefont {Zoller}}]{chang2010cavity}%
  \BibitemOpen
  \bibfield  {author} {\bibinfo {author} {\bibfnamefont {D.~E.}\ \bibnamefont
  {Chang}}, \bibinfo {author} {\bibfnamefont {C.}~\bibnamefont {Regal}},
  \bibinfo {author} {\bibfnamefont {S.}~\bibnamefont {Papp}}, \bibinfo {author}
  {\bibfnamefont {D.}~\bibnamefont {Wilson}}, \bibinfo {author} {\bibfnamefont
  {J.}~\bibnamefont {Ye}}, \bibinfo {author} {\bibfnamefont {O.}~\bibnamefont
  {Painter}}, \bibinfo {author} {\bibfnamefont {H.~J.}\ \bibnamefont
  {Kimble}},\ and\ \bibinfo {author} {\bibfnamefont {P.}~\bibnamefont
  {Zoller}},\ }\bibfield  {title} {\bibinfo {title} {Cavity opto-mechanics
  using an optically levitated nanosphere},\ }\href@noop {} {\bibfield
  {journal} {\bibinfo  {journal} {Proceedings of the National Academy of
  Sciences}\ }\textbf {\bibinfo {volume} {107}},\ \bibinfo {pages} {1005}
  (\bibinfo {year} {2010})}\BibitemShut {NoStop}%
\bibitem [{\citenamefont {Einstein}\ \emph {et~al.}(1905)\citenamefont
  {Einstein} \emph {et~al.}}]{einstein1905electrodynamics}%
  \BibitemOpen
  \bibfield  {author} {\bibinfo {author} {\bibfnamefont {A.}~\bibnamefont
  {Einstein}} \emph {et~al.},\ }\bibfield  {title} {\bibinfo {title} {On the
  electrodynamics of moving bodies},\ }\href@noop {} {\bibfield  {journal}
  {\bibinfo  {journal} {Annalen der physik}\ }\textbf {\bibinfo {volume}
  {17}},\ \bibinfo {pages} {891} (\bibinfo {year} {1905})}\BibitemShut
  {NoStop}%
\bibitem [{\citenamefont {Braginski}\ and\ \citenamefont
  {Manukin}(1967)}]{braginski1967ponderomotive}%
  \BibitemOpen
  \bibfield  {author} {\bibinfo {author} {\bibfnamefont {V.}~\bibnamefont
  {Braginski}}\ and\ \bibinfo {author} {\bibfnamefont {A.}~\bibnamefont
  {Manukin}},\ }\bibfield  {title} {\bibinfo {title} {Ponderomotive effects of
  electromagnetic radiation},\ }\href@noop {} {\bibfield  {journal} {\bibinfo
  {journal} {Sov. Phys. JETP}\ }\textbf {\bibinfo {volume} {25}},\ \bibinfo
  {pages} {653} (\bibinfo {year} {1967})}\BibitemShut {NoStop}%
\bibitem [{\citenamefont {Matsko}\ \emph {et~al.}(1996)\citenamefont {Matsko},
  \citenamefont {Zubova},\ and\ \citenamefont {Vyatchanin}}]{matsko1996value}%
  \BibitemOpen
  \bibfield  {author} {\bibinfo {author} {\bibfnamefont {A.}~\bibnamefont
  {Matsko}}, \bibinfo {author} {\bibfnamefont {E.}~\bibnamefont {Zubova}},\
  and\ \bibinfo {author} {\bibfnamefont {S.}~\bibnamefont {Vyatchanin}},\
  }\bibfield  {title} {\bibinfo {title} {The value of the force of radiative
  friction},\ }\href@noop {} {\bibfield  {journal} {\bibinfo  {journal} {Optics
  communications}\ }\textbf {\bibinfo {volume} {131}},\ \bibinfo {pages} {107}
  (\bibinfo {year} {1996})}\BibitemShut {NoStop}%
\bibitem [{\citenamefont {Hartemann}\ and\ \citenamefont
  {Luhmann~Jr}(1995)}]{hartemann1995classical}%
  \BibitemOpen
  \bibfield  {author} {\bibinfo {author} {\bibfnamefont {F.}~\bibnamefont
  {Hartemann}}\ and\ \bibinfo {author} {\bibfnamefont {N.}~\bibnamefont
  {Luhmann~Jr}},\ }\bibfield  {title} {\bibinfo {title} {Classical
  electrodynamical derivation of the radiation damping force},\ }\href@noop {}
  {\bibfield  {journal} {\bibinfo  {journal} {Physical review letters}\
  }\textbf {\bibinfo {volume} {74}},\ \bibinfo {pages} {1107} (\bibinfo {year}
  {1995})}\BibitemShut {NoStop}%
\bibitem [{\citenamefont {Tamburini}\ \emph {et~al.}(2010)\citenamefont
  {Tamburini}, \citenamefont {Pegoraro}, \citenamefont {Di~Piazza},
  \citenamefont {Keitel},\ and\ \citenamefont
  {Macchi}}]{tamburini2010radiation}%
  \BibitemOpen
  \bibfield  {author} {\bibinfo {author} {\bibfnamefont {M.}~\bibnamefont
  {Tamburini}}, \bibinfo {author} {\bibfnamefont {F.}~\bibnamefont {Pegoraro}},
  \bibinfo {author} {\bibfnamefont {A.}~\bibnamefont {Di~Piazza}}, \bibinfo
  {author} {\bibfnamefont {C.~H.}\ \bibnamefont {Keitel}},\ and\ \bibinfo
  {author} {\bibfnamefont {A.}~\bibnamefont {Macchi}},\ }\bibfield  {title}
  {\bibinfo {title} {Radiation reaction effects on radiation pressure
  acceleration},\ }\href@noop {} {\bibfield  {journal} {\bibinfo  {journal}
  {New Journal of Physics}\ }\textbf {\bibinfo {volume} {12}},\ \bibinfo
  {pages} {123005} (\bibinfo {year} {2010})}\BibitemShut {NoStop}%
\bibitem [{\citenamefont {Singal}(2016)}]{singal2016compatibility}%
  \BibitemOpen
  \bibfield  {author} {\bibinfo {author} {\bibfnamefont {A.~K.}\ \bibnamefont
  {Singal}},\ }\bibfield  {title} {\bibinfo {title} {Compatibility of
  larmor’s formula with radiation reaction for an accelerated charge},\
  }\href@noop {} {\bibfield  {journal} {\bibinfo  {journal} {Foundations of
  Physics}\ }\textbf {\bibinfo {volume} {46}},\ \bibinfo {pages} {554}
  (\bibinfo {year} {2016})}\BibitemShut {NoStop}%
\bibitem [{\citenamefont {Wineland}\ and\ \citenamefont
  {Itano}(1979)}]{wineland1979laser}%
  \BibitemOpen
  \bibfield  {author} {\bibinfo {author} {\bibfnamefont {D.~J.}\ \bibnamefont
  {Wineland}}\ and\ \bibinfo {author} {\bibfnamefont {W.~M.}\ \bibnamefont
  {Itano}},\ }\bibfield  {title} {\bibinfo {title} {Laser cooling of atoms},\
  }\href@noop {} {\bibfield  {journal} {\bibinfo  {journal} {Physical Review
  A}\ }\textbf {\bibinfo {volume} {20}},\ \bibinfo {pages} {1521} (\bibinfo
  {year} {1979})}\BibitemShut {NoStop}%
\bibitem [{\citenamefont {Wu}\ \emph {et~al.}(2013)\citenamefont {Wu},
  \citenamefont {Horsley}, \citenamefont {Artoni},\ and\ \citenamefont
  {La~Rocca}}]{wu2013radiation}%
  \BibitemOpen
  \bibfield  {author} {\bibinfo {author} {\bibfnamefont {J.-H.}\ \bibnamefont
  {Wu}}, \bibinfo {author} {\bibfnamefont {S.}~\bibnamefont {Horsley}},
  \bibinfo {author} {\bibfnamefont {M.}~\bibnamefont {Artoni}},\ and\ \bibinfo
  {author} {\bibfnamefont {G.~C.}\ \bibnamefont {La~Rocca}},\ }\bibfield
  {title} {\bibinfo {title} {Radiation damping optical enhancement in cold
  atoms},\ }\href@noop {} {\bibfield  {journal} {\bibinfo  {journal} {Light:
  Science \& Applications}\ }\textbf {\bibinfo {volume} {2}},\ \bibinfo {pages}
  {e54} (\bibinfo {year} {2013})}\BibitemShut {NoStop}%
\bibitem [{\citenamefont {Horsley}\ \emph {et~al.}(2011)\citenamefont
  {Horsley}, \citenamefont {Artoni},\ and\ \citenamefont
  {La~Rocca}}]{horsley2011radiation}%
  \BibitemOpen
  \bibfield  {author} {\bibinfo {author} {\bibfnamefont {S.}~\bibnamefont
  {Horsley}}, \bibinfo {author} {\bibfnamefont {M.}~\bibnamefont {Artoni}},\
  and\ \bibinfo {author} {\bibfnamefont {G.}~\bibnamefont {La~Rocca}},\
  }\bibfield  {title} {\bibinfo {title} {Radiation damping in atomic photonic
  crystals},\ }\href@noop {} {\bibfield  {journal} {\bibinfo  {journal}
  {Physical Review Letters}\ }\textbf {\bibinfo {volume} {107}},\ \bibinfo
  {pages} {043602} (\bibinfo {year} {2011})}\BibitemShut {NoStop}%
\bibitem [{\citenamefont {Gould}(1997)}]{gould1997laser}%
  \BibitemOpen
  \bibfield  {author} {\bibinfo {author} {\bibfnamefont {P.}~\bibnamefont
  {Gould}},\ }\bibfield  {title} {\bibinfo {title} {Laser cooling of atoms to
  the doppler limit},\ }\href@noop {} {\bibfield  {journal} {\bibinfo
  {journal} {American Journal of Physics}\ }\textbf {\bibinfo {volume} {65}},\
  \bibinfo {pages} {1120} (\bibinfo {year} {1997})}\BibitemShut {NoStop}%
\bibitem [{\citenamefont {Jackson}(2021)}]{jackson2021classical}%
  \BibitemOpen
  \bibfield  {author} {\bibinfo {author} {\bibfnamefont {J.~D.}\ \bibnamefont
  {Jackson}},\ }\href@noop {} {\emph {\bibinfo {title} {Classical
  electrodynamics}}}\ (\bibinfo  {publisher} {John Wiley \& Sons},\ \bibinfo
  {year} {2021})\BibitemShut {NoStop}%
\bibitem [{Note1()}]{Note1}%
  \BibitemOpen
  \bibinfo {note} {The rest frame of the particle is the inertial frame moving
  with velocity $\protect \mathbf {v}$ with respect to the laboratory frame. We
  use prime notation to denote quantities in the rest frame.}\BibitemShut
  {Stop}%
\bibitem [{\citenamefont {Chaumet}\ and\ \citenamefont
  {Nieto-Vesperinas}(2000)}]{chaumet2000time}%
  \BibitemOpen
  \bibfield  {author} {\bibinfo {author} {\bibfnamefont {P.~C.}\ \bibnamefont
  {Chaumet}}\ and\ \bibinfo {author} {\bibfnamefont {M.}~\bibnamefont
  {Nieto-Vesperinas}},\ }\bibfield  {title} {\bibinfo {title} {Time-averaged
  total force on a dipolar sphere in an electromagnetic field},\ }\href@noop {}
  {\bibfield  {journal} {\bibinfo  {journal} {Optics letters}\ }\textbf
  {\bibinfo {volume} {25}},\ \bibinfo {pages} {1065} (\bibinfo {year}
  {2000})}\BibitemShut {NoStop}%
\bibitem [{\citenamefont {Abbassi}\ and\ \citenamefont
  {Mehrany}(2018)}]{abbassi2018inclusion}%
  \BibitemOpen
  \bibfield  {author} {\bibinfo {author} {\bibfnamefont {M.~A.}\ \bibnamefont
  {Abbassi}}\ and\ \bibinfo {author} {\bibfnamefont {K.}~\bibnamefont
  {Mehrany}},\ }\bibfield  {title} {\bibinfo {title} {Inclusion of the
  backaction term in the total optical force exerted upon rayleigh particles in
  nonresonant structures},\ }\href@noop {} {\bibfield  {journal} {\bibinfo
  {journal} {Physical Review A}\ }\textbf {\bibinfo {volume} {98}},\ \bibinfo
  {pages} {013806} (\bibinfo {year} {2018})}\BibitemShut {NoStop}%
\bibitem [{\citenamefont {Abbassi}\ and\ \citenamefont
  {Mehrany}(2022)}]{abbassi2022self}%
  \BibitemOpen
  \bibfield  {author} {\bibinfo {author} {\bibfnamefont {M.~A.}\ \bibnamefont
  {Abbassi}}\ and\ \bibinfo {author} {\bibfnamefont {K.}~\bibnamefont
  {Mehrany}},\ }\bibfield  {title} {\bibinfo {title} {Self-induced backaction
  in optical waveguides},\ }\href@noop {} {\bibfield  {journal} {\bibinfo
  {journal} {Optics Express}\ }\textbf {\bibinfo {volume} {30}},\ \bibinfo
  {pages} {42967} (\bibinfo {year} {2022})}\BibitemShut {NoStop}%
\bibitem [{\citenamefont {Abbassi}\ and\ \citenamefont
  {Mehrany}(2019)}]{abbassi2019green}%
  \BibitemOpen
  \bibfield  {author} {\bibinfo {author} {\bibfnamefont {M.~A.}\ \bibnamefont
  {Abbassi}}\ and\ \bibinfo {author} {\bibfnamefont {K.}~\bibnamefont
  {Mehrany}},\ }\bibfield  {title} {\bibinfo {title} {Green's-function
  formulation for studying the backaction cooling of a levitated nanosphere in
  an arbitrary structure},\ }\href@noop {} {\bibfield  {journal} {\bibinfo
  {journal} {Physical Review A}\ }\textbf {\bibinfo {volume} {100}},\ \bibinfo
  {pages} {023823} (\bibinfo {year} {2019})}\BibitemShut {NoStop}%
\bibitem [{\citenamefont {Novotny}\ and\ \citenamefont
  {Hecht}(2012)}]{novotny2012principles}%
  \BibitemOpen
  \bibfield  {author} {\bibinfo {author} {\bibfnamefont {L.}~\bibnamefont
  {Novotny}}\ and\ \bibinfo {author} {\bibfnamefont {B.}~\bibnamefont
  {Hecht}},\ }\href@noop {} {\emph {\bibinfo {title} {Principles of
  nano-optics}}}\ (\bibinfo  {publisher} {Cambridge university press},\
  \bibinfo {year} {2012})\BibitemShut {NoStop}%
\bibitem [{\citenamefont {Zangwill}(2013)}]{zangwill2013modern}%
  \BibitemOpen
  \bibfield  {author} {\bibinfo {author} {\bibfnamefont {A.}~\bibnamefont
  {Zangwill}},\ }\href@noop {} {\emph {\bibinfo {title} {Modern
  electrodynamics}}}\ (\bibinfo  {publisher} {Cambridge University Press},\
  \bibinfo {year} {2013})\BibitemShut {NoStop}%
\bibitem [{\citenamefont {Ritz}\ and\ \citenamefont
  {Einstein}(1909)}]{ritz1909gegenwartigen}%
  \BibitemOpen
  \bibfield  {author} {\bibinfo {author} {\bibfnamefont {W.}~\bibnamefont
  {Ritz}}\ and\ \bibinfo {author} {\bibfnamefont {A.}~\bibnamefont
  {Einstein}},\ }\bibfield  {title} {\bibinfo {title} {Zum gegenw{\"a}rtigen
  stand des strahlungsproblems},\ }\href@noop {} {\bibfield  {journal}
  {\bibinfo  {journal} {Physikalische Zeitschrift}\ }\textbf {\bibinfo {volume}
  {10}},\ \bibinfo {pages} {323} (\bibinfo {year} {1909})}\BibitemShut
  {NoStop}%
\end{thebibliography}
%

\end{document}